\documentclass{article}

\usepackage{arxiv}

\usepackage[utf8]{inputenc} 
\usepackage[T1]{fontenc}    
\usepackage{hyperref}       
\usepackage{url}            
\usepackage{booktabs}       
\usepackage{amsfonts} 
\usepackage{graphicx} 
\usepackage{nicefrac}       
\usepackage{microtype}      
\usepackage{lipsum}
\usepackage{amsmath,amssymb,amsthm}
\theoremstyle{definition}
\DeclareMathOperator{\E}{\mathbb{E}}
\DeclareMathOperator{\Prob}{\mathbb{P}}

\newtheorem{theorem}{Theorem}[section]
\newtheorem{lemma}{Lemma}[section]
\newtheorem{definition}{Definition}[section]
\newtheorem{example}{Example}[section]

\title{Projection Pursuit with applications to scRNA sequencing data}

\author{
	 Elvis Han Cui \\
	Department of Biostatistics\\
	University of California, Los Angeles\\
	\texttt{elviscuihan@g.ucla.edu} \\
	\And
 Heather Zhou\\
 Department of Statistics\\
  University of California, Los Angeles\\
  \texttt{heatherjzhou@ucla.edu} \\
}

\begin{document}
\maketitle

\begin{abstract}
In this paper, we explore the limitations of PCA as a dimension reduction technique and study its extension, projection pursuit (PP), which is a broad class of linear dimension reduction methods. We first discuss the relevant concepts and theorems and then apply PCA and PP (with negative standardized Shannon's entropy as the projection index) on single cell RNA sequencing data.

\end{abstract}

\section{PCA and its issues}
PCA is a popular dimension reduction technique commonly applied to scRNA sequencing data. There are several ways of deriving PCA, such as Karhunen-Loeve transform, Hotelling transform, and minimizing square error (chapter 7 and 16 of \cite{izenman2008}). Despite of huge success in practice, we will illustrate three drawbacks of PCA. Due to these drawbacks, it is reasonable for us to seek alternatives of PCA.

\subsection{Asymptotic distributions of eigenvalues}
It is well known that the eigenvalues of sample covariance matrix is not consistent in high dimensional cases. Suppose we observe $\mathbf{X}=(x_1,\cdots,x_n)^T,x_i\in\mathbb{R}^d$. If both $n,d\rightarrow\infty$, then we have the famous \textbf{quarter-circle law} (a.k.a. Marcenko-Pastur) in statistical physics:

\begin{theorem}[Marcenko-Pastur Law]
	Suppose $\mathbf{X^TX}\sim\mathcal{W}_d(n,\mathbf{I}_d)$. Define the \textbf{empirical spectral distribution}
	$$G_{d}(k)=\frac{1}{d}\#\{\hat{\lambda}_j\le k\}$$
	Where $\hat{\lambda}_j$'s are eigen-values of $\frac{\mathbf{X^TX}}{n}$. If $\frac{d}{n}\rightarrow \gamma\in(0,\infty)$, then, $G_d(k)\rightarrow G(k)$ almost surely, where the limiting distribution $G(k)$ has density $g(k)=G'(k)$:
	$$g(k)=\frac{\sqrt{(b_+-k)(k-b_-)}}{2\pi\gamma k},b_\pm=(1\pm\sqrt{\gamma})^2$$
\end{theorem}

Therefore, in high dimensions, eigenvalues and eigen-vectors of sample covariance matrix are not consistent and PCA fails naturely.

\subsection{Components other than uncorrelatedness}
Every principal component is uncorrelated with each other but not independent. What if we want independence ? Thus, we should search for other criterions other than "maximizing variance". It turns out that "mutual information" in information theory will be a perfect alternative as we will illustrate in nect section.

Besides, if we want sparsity on the support of eigen-vectors, then we may consider the so-called "sparse PCA" (\cite{wainwright2019}). If we want our estimation to be more robust, then we may consider robust PCA or other robust estimations (\cite{huber1985}). Moreover, what if we are facing a supervised learning problem instead of just dimension reduction ? All in a word, there is a beatiful framework that unifies all these aspects known as \textbf{projection pursuit}. We will study it briefly in next section.

\subsection{PCA is not suitable for clustering}

It is worth noting that PCA is \textbf{NOT} designed for clustering, given its objective function. Therefore, it is not surprise that PCA behaves poorly in some situations for clustering. The following figure shows such situation (left: original data, middle: PCA, right: PP).

\begin{figure}[!ht]
	\begin{minipage}{0.3\textwidth}
		\centering
		\includegraphics[width=1\linewidth]{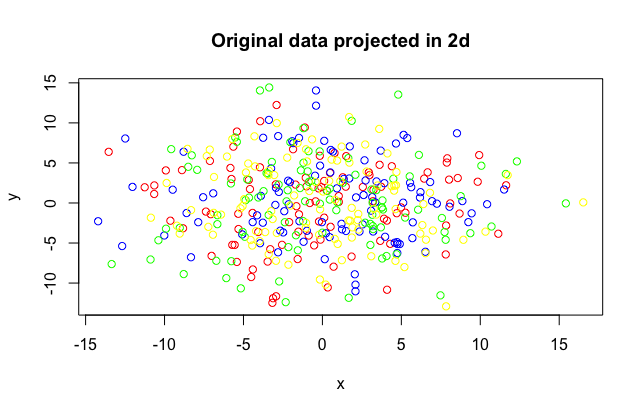}
	\end{minipage}
	\begin{minipage}{0.3\textwidth}
		\centering
		\includegraphics[width=1\linewidth]{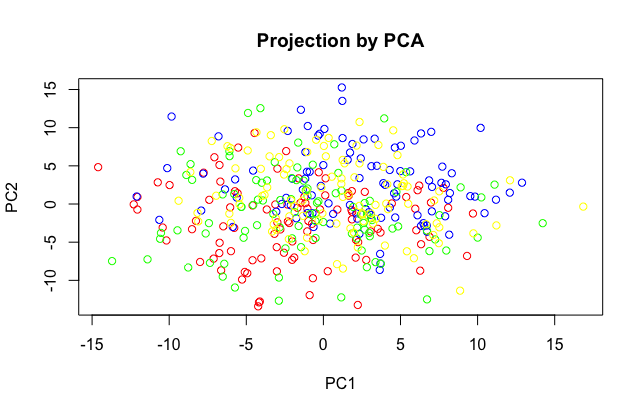}
	\end{minipage}
	\begin{minipage}{0.3\textwidth}
		\centering
		\includegraphics[width=1\linewidth]{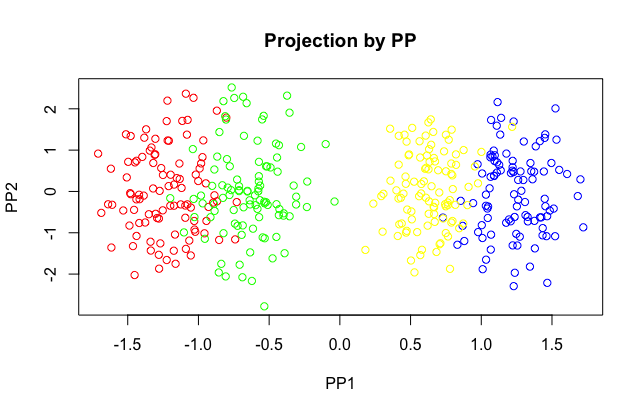}
	\end{minipage}
\end{figure}

\section{Projection pursuit: basic concepts and properties}
There are many non-linear alternatives for PCA, such as SNE, t-SNE ( \cite{jessica2019a}), kernel PCA,  principal curves and surfaces (\cite{izenman2008} and \cite{esl}), etc. However, linear dimension reduction is still the most important since  \textbf{Fourier inversion theorem} (\cite{dabrowska2019} and \cite{jessica2019b}) tells us that the distribution of any random vector is uniquely determined by its 1-d projection.

More generally speaking, suppose we are ineterested in a "lower dimensional projection" (not just 1-d) of our high dimensional data. Here "interesting" refers to different kinds of criterions. For instance, if we are interested in finding a direction such that the variance of projection is maximized, then we get ordinary PCA. Also, such lower dimensional projection should bypass "the curse of dimensionality" since there are few points in a high dimensional space. Besides, rubustness and computational efficiency should be taken into account. All these ideas can be found in Huber (\cite{huber1985}) and lead us to the definition of \textbf{projection pursuit}.

\subsection{Definition}
First, we need a "loss function" or "index" that measures how well our projection is, this is called \textbf{projection index} in literature.
\begin{definition}[Projection index \cite{friedman1974}] Let $X$ be a random vector and $A$ is a matrix. A \textbf{projection index} is a functional $Q:F_A\to\mathbb{R}$ where $F_A$ is the distribution of $Z=AX$. We will denote projection index as
	$Q(Z)$ or $Q(F_A)$.
\end{definition}

Second, given a specific projection index $Q$ and data $\mathbf{X}$, we need to find a direction that maximizes such index. This is known as \textbf{projection pursuit} in literature. 
\begin{definition}[Projection pursuit (PP) \cite{huber1985}]
	\textbf{Projection pursuit} (PP) searches for a projection $A$ maximizing (or minimizing) a projection index $Q(Z)$. Usually, the projection will be less or equal to 3 dimensions for visualization.
\end{definition}
\subsection{Diaconis-Freedman theorem}
\begin{theorem}[Diaconis and Freedman, 1984; Bickel et al, 2018]
Suppose $x_j$'s are i.i.d. and the projection $z$ was assigned the uniform distribution in $\mathbb{S}^{d-1}$, then as $d,n\rightarrow\infty$, $\forall\epsilon>0$
$$\mathbb{P}[\rho(\widehat{G}_z,\Phi)<\epsilon]\rightarrow1$$
where $\rho$ is the Levy-Prohorov metric:
$$\rho(\mu,\nu):=\inf\{\epsilon>0|\mu(A)\le\nu(A^\epsilon)+\epsilon\text{ and }\nu(A)\le\mu(A^\epsilon)+\epsilon\ \forall A\in\mathcal{B}(\Omega)\}$$
Thus, non-Gaussian projections would indeed be rare and interesting.
\end{theorem}

\subsection{Bickel-Kur-Nadler theorem}
A natural question on the properties of PP is: given an arbitrary cumulative distribution function $G$ (i.e. mixtures of multivariate Gaussians), how well can our projected data approximates such distribution? Such question is answered elegantly by Bickel, Kur and Nadler.
\begin{theorem}[Bickel-Kur-Nadler \cite{bickel2018}]
	Suppose $\frac{d}{n}\rightarrow\infty$. Let $G(t)$ be an \textbf{arbitrary cumulative distribution function} with mean 0. There there $\exists$ a sequence of projections $z=z_n\in\mathbb{S}^{d-1}$ s.t. the following holds:
	$$\lim_{n\rightarrow\infty}\lVert\widehat{G}_Z-G\lVert_\infty=0$$
	That is, $\widehat{G}_Z$ converges uniformly (thus, weakly) to $G$.
\end{theorem}
Basically, this theorem says: if different types of cells have different gene expression, they there \textbf{EXISTs} a projection pursuit program s.t. we could visualize high dimensional clustered data in one or two or three dimensions. 

\section{Examples of projection indexes} In this section, we present some examples of various projection indexes, among which some are original. Those indexes can be classified as 2 types: non-entropy based and entropy based indexes. The former includes a blanket of classical statistical methods such as Fisher's LDA, PCA and canonical correlation analysis. The latter has strong connection with another technique called \textbf{independent component analysis} (ICA).

\subsection{Non-entropy based indexes}
\begin{example}[Sample mean \cite{huber1985}]
	Take $Q(a^TX)=\E(a^TX)=a^T\mu\text{ s.t. }\lVert a\lVert_2=1$. This is a 1-d projection and a natural estimation is $\widehat{Q}(a^TX)=\frac{1}{n}\sum_{i=1}^na^Tx_i$. Using Lagragian multiplier, this index is maximized by $a_0=\frac{\mu}{\lVert\mu\lVert}$. Thus, sample mean can be derived via PP as $a_0Q(a_0^TX)$.
\end{example}

\begin{example}[Principal component analysis] Let $Q(a^TX)=Var(a^TX)\text{ s.t. }\lVert a\lVert_2=1$. Then we get first principal component. Next, take $Q(b^TX)=Var(b^TX)\text{ s.t. }\lVert b\lVert_2=1, \langle a,b\rangle=0$. We get second principal component.
\end{example}

\begin{example}[Canonical correlation analysis (CCA)]
Suppose $X\in\mathbb{R}^{d_1}$ and $Y\in\mathbb{R}^{d_2}$, then 
$$Q(a^TX,b^TY)=Corr(a^TX,b^TY)$$
corresponds to CCA. Note that CCA is affine invariant, that is , if we take $Z=(a^X,b^TY)^T$, then $Q(Z)=Q(t_1Z+t_2)$. This is known as \textbf{class III} index in \cite{huber1985}.
\end{example}
\begin{example}[Fisher's linear discriminant analysis (LDA)]
	Suppose both $X,Y\in\mathbb{R}^d$ and they share the same covariance matrix $Var(X)=Var(Y)=\Sigma$, then
	$Q(a^TX,a^TY)=\frac{(a^T\E(X)-a^T\E(Y))^2}{a^T\Sigma a}$
	corresponds to the famous Fisher's linear discriminant function. To generalize this to multi-dimensional cases, we could take
	$$Q(AX,AY)=\frac{\lVert A\E(X)-A\E(Y)\lVert_2^2}{Tr(A\Sigma A)}\text{ s.t. }A^TA=I$$
	Both trace and $L_2$ norm can be replaced by other suitable measurements (i.e. Frobenious norm) and $\det$ corresponds to multiple discriminant analysis (MDA). Note that the constraint $A^TA=I$ can be solved alternatively. For instance, solving 1-d LDA gives us
	$$a=\frac{\Sigma^{-1}(\mu_1-\mu_2)}{\lVert\Sigma^{-1}(\mu_1-\mu_2)\lVert_2}$$
\end{example}

\begin{example}[Johnson-Lindenstrauss embedding \cite{wainwright2019}]
	Supoose we want to preserve inner product or Euclidean distance among points after projection, we could take 
	$$Q(AX)=Q(Z)=\sum_{i,j}\left|\lVert z_i-z_j\lVert_2^2-\lVert x_i-x_j\lVert_2^2\right|$$
	And note that in an inner product space, we have $4\langle x,y\rangle=\lVert x+y\lVert^2-\lVert x-y\lVert^2$ where $\lVert\cdot\lVert$ is any norm (for details, see \cite{dabrowska2019}). Therefore, if distance induced by norm is preserved, then inner product is preserved automatically. However, maximizing this projection index requires solving a non-linear system which is computationally expansive (when both $n$ and $d$ is large, this is impossible). Therefore, we resort to a weaker constraint: "the distance/inner product among points are \textbf{almostly} preserved". 
	
	More precisely, given a tolerance level $\delta$ and a confidence level $1-\epsilon$, we want to find a linear projection s.t. $$(1-\delta)\lVert x_i-x_j\lVert_2^2\le\lVert z_i-z_j\lVert_2^2\le(1+\delta)\lVert x_i-x_j\lVert_2^2 \text{ holds w.p. } 1-\epsilon$$
	Johnson-Lindenstrauss embedding provides a perfect solution to this set-up: Let $A\in\mathbb{R}^{r\times d}$ be filled with sub-Gaussian elements (i.e. normal, Rademacher, etc.), then $A$ is the linear projection we want. For more details of this embedding, see chapter 2 of \cite{wainwright2019}.
\end{example}

\begin{example}[Linear SNE] SNE originates from \cite{hinton2002} which is a non-linear projection and non-convex problem. However, if we force the lower dimensional representation to be a linear projection of original data, then we can get the projection index
	\begin{align*}
		Q(AX)=-&\sum_{i=1}^n\sum_{j=1}^n p_{ij}\log(q_{ij})\\
		p_{ij}=\frac{\phi(\frac{\lVert x_i-x_j\lVert_2}{\sigma_i})}{\sum_{k\not= i}\phi(\frac{\lVert x_i-x_k\lVert_2}{\sigma_i})}
		 ,&\ q_{ij}=\frac{\phi(\lVert Ax_i-Ax_j\lVert_2)}{\sum_{k\not= i}\phi(\lVert Ax_i-Ax_k\lVert_2)}
	\end{align*}
	Where $\phi(\cdot)$ denotes the pdf of a standard normal random variable. Here $p_{ij}$'s are constants and this is a weighted negative sum of $\log q_{ij}$. Note that the denominator of $p_{ij}$ is not involved in optimization, Moreover, if we replace $\sum_{j=1}^n$ by integration, then we do not need to normalize $q_{ij}$. The resulting projection index corresponds to "finding a projection that is mostly similar to high dimensional Gaussian distribution", which is the opposite of entropy-based indexes discussed in nect subsection.
\end{example}
\begin{example}[Liquid association \cite{jessica2019a}]
	Taking $Q(a^TX,b^TY,c^TZ)=\E(a^TXb^TYc^TZ)$ gives us a combination of PP and liquid association. For more details, see \cite{jessica2019a}.
\end{example}
\begin{example}[Artifial neural networks \cite{izenman2008}] PP for regression has a beatiful connection with Komogorov's universal approximation theorem: any continuous function with a compact support in $\mathbb{R}^d$ can be approximated by the following conditional expectation $\E(Y|X)$. Let $Y$, a single output variable and $X$, a random vector in $\mathbb{R}^d$ be modeled as
	\begin{align*}
		Y&=a_0+\sum_{j=1}^tf_j(\beta_{0j}+X^T\beta_j)+\epsilon\\
	\E(\epsilon)&=0,Var(\epsilon)=\sigma^2
	\end{align*}
For instance, if $\E(Y|X)=X_1X_2$, then we can rewrite $X_1X_2=\frac{1}{4}(X_1+X_2)^2+\frac{1}{4}(X_1-X_2)^2$, a linear combination of projections of $(X_1,X_2)$. In the language of statistics, $f_j(\cdot)$ is called \textbf{ridge function} while \textbf{activation function} is used in the field of machine learning. In fact, $\E(Y|X)$ has the same form as a two-layer artifial neural network. Then, define our projection index to be
$$Q(\beta^TX+\beta_0)=\sum_{i=1}^n\left\{y_i-a_0+\sum_{j=1}^tf_j(\beta_{0j}+x_i^T\beta_j)\right\}^2$$
Where $\beta=(\beta_1,\cdots,\beta_d)$ and $\beta_0=(\beta_{01},\cdots,\beta_{0d})^T$. Such index can be maximized by the so-called \textbf{back-propagation} algorithms. For more details, see \cite{friedman1987}, \cite{icabook} and \cite{izenman2008}.
\end{example}
\begin{example}[Density approximation using Hellinger distance \cite{huber1985}]
	Suppose we are interested in approximating the density function $f$ in $\mathbb{R}^d$, then we could use \textbf{multiplicative decompositions} (\cite{huber1985})
	$$f_k(x)=f_0(x)\prod_{i=1}^k h_i(a_i^T X)$$
	to approximate $f$ where $f_0$ is a standard density in $\mathbb{R}^d$ (e.g. multivariate Gaussian). To measure the distance of $f_k$ and $f$, we need a metric in the space of probability densities, such metric can be taken as \textbf{Hellinger distance}:
	$$Q(AX)=\text{Hellinger}(f,f_k)=\int(\sqrt{f}-\sqrt{f_k})^2dx$$
\end{example}

\subsection{Entropy based indexes}
Although a lot of linear models can be viewed as PP, most literatures concerned about \textbf{entropy based indexes} and their approximations. The reasons are:
\begin{itemize}
    \item Due to Diaconis-Freedman theorem, non-Gaussian projections would indeed be rare and interesting.
    \item There is a strong connection between PP and independent component analysis (ICA) where the latter uses information theory as its theoretical foundations.
    \item In many unsupervised applications, PCA is suffice to give good visualizations (though in theory it may not perform good).
\end{itemize}

\subsubsection{Entropy and its properties}
To get into more technical details, we need some basic concepts from information theory. For details, see \textbf{appendix} and \cite{cover2006}.

Applications of information theory to PP (based two lemmas on entropy in appendix) are Fisher's information and mutual information.
\begin{example}[Fisher's information]
	Let $X$ have continuous differentiable density $f$ w.r.t. $\lambda_d$, then the Fisher's information projection index is
	$$Q(X)=\sigma^2\int(\frac{f'}{f})^2fd\lambda_d-1$$
	Or equivalently,
	$$Q(X)=\sigma^2\int(\frac{f'}{f}-\frac{\phi '}{\phi})^2fd\lambda_d$$
	where $\sigma^2=Var(X)$ and $\phi(\cdot)$, as usual, the pdf of $\mathcal{N}(0,1)$. Then this index is affine invariant due to lemma \ref{entropyaffine}.
\end{example}
\begin{example}[Mutual information (MI) \cite{komogorov1956},\cite{cover2006}]
	The general definition of mutual information is given in \cite{komogorov1956}. The intuition of MI is "the amount of information contained in $X$ about $Y$". Let $(\Omega,\mathcal{F},\mathbb{P})$ be a probability space and $X,Y$ are a pair of $\mathcal{F}/\mathcal{B}(\mathbb{R})$-measurable random variables. Then the \textbf{mutual information} of $X$ and $Y$ is
	$$I(X,Y)=\int_{\mathbb{R}^2}\Prob_{X,Y}(dxdy)\log\frac{\Prob_{X,Y}(dxdy)}{\Prob_X(dx)\Prob_Y(dy)}$$
	where $\Prob_{X,Y}$ is the probability measure induced by $(X,Y)$. Note that, $I(X,Y)\ge 0$ and equality is attained iff $X$ and $Y$ are independent. Thus, mutual information can be used as a projection index for measuring indepence. Now suppose  $X$ is random vector and $A\in\mathbb{R}^{d\times d}$ and define
	$$Q(AX)=-\int_{\mathbb{R}^d}\Prob_{AX}(dx_1\cdots dx_d)\log\frac{\Prob_{AX}(dx_1\cdots dx_d)}{\Prob_{a_1^TX}(dx_1)\cdots \Prob_{a_d^TX}(dx_d)}$$
	Then $Q(AX)\le 0$ with equality attained iff all projected directions are independent. Note that this is different from PCA since the latter only gives uncorrelated projections.
\end{example}

Equipped with new weapons, we will introduce a frequently used entroy based index: stadardized negative Shannon's entropy . Then we will discuss how to approximate it in practice. Methods based on cumulant and non-polynomial functions will be introduced.

\begin{example}[Stadardized negative Shannon's entropy]
	Suppose we have some prior knowledges about our data (i.e. $\E(X)=\mu,\E(\sin(\beta^TX))=c$) and we want our data is away from "chaos", which is measured by \textbf{entropy} in information theory, then due to the following theorem, we would expect our projected data to be as far away from a family of particular distributions as possible. Therefore, we choose our projection index to be
	\begin{align*}
	Q(a^TX)&=-\int\log(\frac{\phi}{f})fdx
	\end{align*}
	where $\phi(\cdot)$, the pdf of $\mathcal{N}(0,1)$, $f$, the pdf of projected data and $\sigma=\sqrt{Var(a^TX)}$. One reason we want to set $f$ to be far away from Gaussian is Diaconis-Freedman and another is the following.
	\begin{theorem}[Maximum entropy principle \cite{bickedoskum2001},\cite{sundaberg2019},\cite{cover2006}] Suppose $X$ is a random vector with density $f$ w.r.t. counting or Lebesgue measure and we have the following constraints:
		$$\int_{\mathbb{R}^d} f(x)\mu(dx)=1,\int_{\mathbb{R}^d}f(x)r_j(x)\mu(dx)=\alpha_j,1\le j\le k $$
		Then $f$ has the following form (w.r.t. $\mu$)
		$$f(x)=A\exp\{\sum_{j=1}^k\eta_jr_j(x)\}$$
		where $A$ and $\eta_j$ are constants s.t. $f$ is a probability density. If we take $k=2,r_1(x)=x\text{ and }r_2(x)=x^2$, then 
		$$f(x)\propto \exp\{\eta_1x+\eta_2x^2\}$$
		which is the density of a univariate Gaussian distribution.
	\end{theorem}
However, we cannot evaluate $\int f\log fdx$ directly since this is an integration. Also, estimation of density $f$ is difficult (kernel estimators will be very poor prone \cite{icabook}). Thus, we have to approximate the entropy directly by some numerical methods. This leads us to the following 2 subsubsections.
\end{example}

\subsubsection{Cumulant-based approximations}
The idea is to use high-order cumulants based on the Hermite polynomials, a complete orthogonalpolynomial basis in $L_2(\Prob)$ where $\Prob$ denotes the standard Gaussian distribution. The are given by $H_0(x)=1$ and
$$H_n(x)=(-1)^n\frac{\phi^{(n)}}{\phi}(x)\text{ for }n\ge 1$$
$$\int\phi(x)H_n(x)H_m(x)dx=n!\delta_{nm}$$
where $\delta_{nm}$ is Kronecker delta. For technical details, see chapter 8 of \cite{dabrowska2019}. If we cut off this expansion at the first 2 non-constants, then we get an approximation of the density function $f(x)$. Then the standardized negative Shannon's entropy of $a^TX$ can be approximated by (e.g. \cite{icabook})
$$\widehat{Q}(a^TX)=\widehat{Q}(Z)=\frac{(\kappa_3(Z))^2}{12}+\frac{(\kappa_4(Z))^2}{48}$$
where $\kappa_3$ and $\kappa_4$ are the usual cumulants. For cumulants in high dimensions, see \cite{mardia1970}.
\subsubsection{Approximation based on non-polynomial functions}
Cumulants are not robust to outliers so we need some other approximations. One mostly used  approximation is based on non-polynomial functions (\cite{comon1994},\cite{friedman1987},\cite{icabook},\cite{izenman2008},\cite{esl}). Satisfying a group of constraints (e.g. section 5.6 of \cite{icabook}, section 15.3 of \cite{izenman2008}), we can approximate $f(x)$ by
$$\widehat{f}(x)=\phi(x)(1+\sum_{i=1}^nc_iG_i(x))$$
Where $c_i=\E(G_i(x))$ and $G_i$'s are non-polynomial functions. Then an approximation of projection index is
$$\widehat{Q}(a^TX)=\widehat{Q}(Z)=\frac{1}{2}\sum_{i=1}^n\left(\E(G_i(Z))\right)^2$$
And we replace $\E$ by sample mean or other common estimators. In section 4, we will let n=1 and choose $G(x)=\frac{1}{\alpha}\log\cosh(\alpha x)$ to approximate standardized negative Shannon's netropy.

\section{Application to scRNA-seq data}

\subsection{Data pre-processing}

We start with the raw UMI count data from PBMC experiments \cite{scRNAseqDataPaper} (data downloaded from Broad Institute Single Cell Portal). First, we separate the data from the first PBMC experiment into 7 matrices (cell by gene) based on the sequencing method. Second, we filter out genes with zero expression in more than 80\% cells. Lastly, we quantile normalize each count matrix so that in each count matrix, the genes have the same empirical distribution in every cell. The dimensions of the resulting count matrices are as follows:

\begin{enumerate}
    \item pbmc1\_CEL\_Seq2: $253\times3,576$
    \item pbmc1\_10xChromiumv2A: $3,222\times623$
    \item pbmc1\_10xChromiumv2B: $3,222\times736$
    \item pbmc1\_10xChromiumv3: $3,222\times1,696$
    \item pbmc1\_Drop\_seq: $3,222\times516$
    \item pbmc1\_Seq\_Well: $3,222\times337$
    \item pbmc1\_inDrops: $3,222\times212$
\end{enumerate}

In addition, we extract the true cell type labels of the cells from meta.txt on the Single Cell Portal, which are derived using marker genes by \cite{scRNAseqDataPaper}. We made sure that the cells in the count matrices matched with the cells in meta.txt.

\subsection{Results}
\begin{figure*}[htbp!]
    \vspace{-0.5in}
    \centerline{\includegraphics[width=0.8\textwidth]{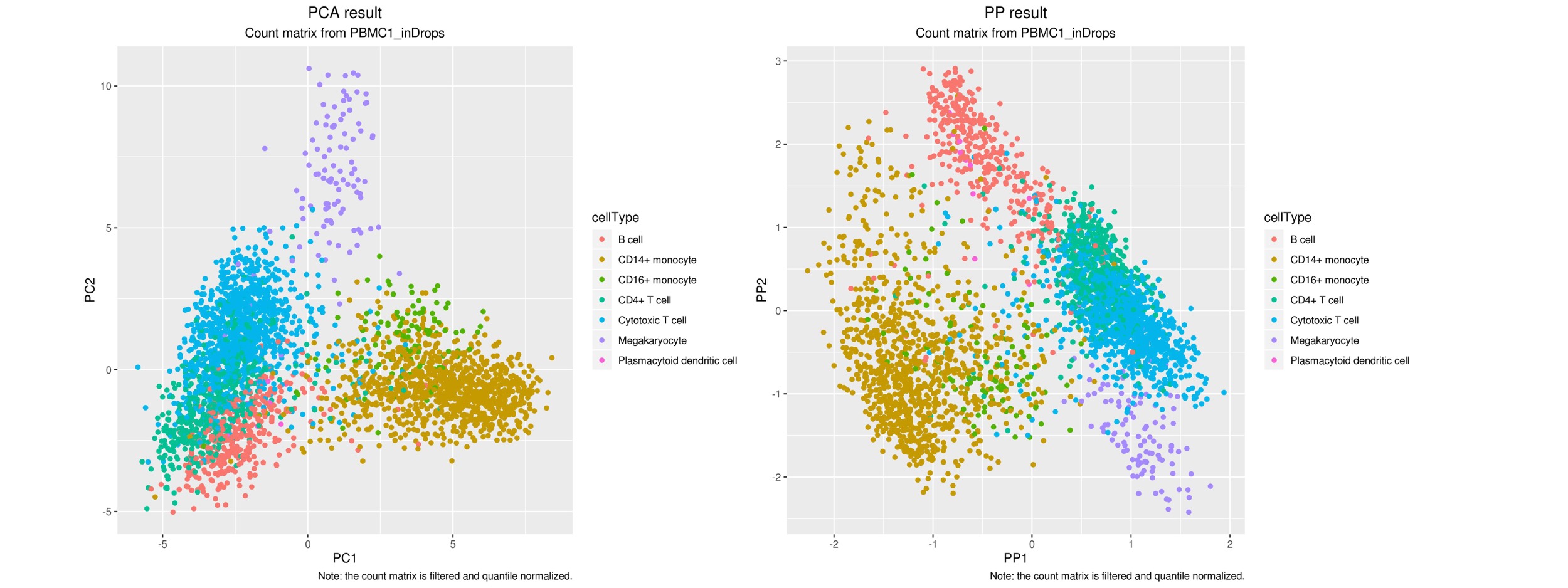}}
    \caption{PCA (left) and PP (right) results on the count matrices from 7 different sequencing methods}
    \label{fig:results}
\end{figure*}

For each of the 7 pre-processed count matrices, we apply PCA and PP (with negative standardized Shannon's entropy as the projection index) and plot the results in Figure~\ref{fig:results}. Each row corresponds to one count matrix from one particular sequencing method. The left column is results from PCA, and the right column is results from PP. Each point in a plot is a cell, and the points are colored based on the true cell type.

We observe from Figure~\ref{fig:results} that PCA actually produces better results than PP in terms of keeping cells of the same cell type close together. For example, for the count matrix from 10x Chromium V2-B (third row of Figure~\ref{fig:results}), the data points of the same color tend to stay close together in the PCA plot (left) more than they do in the PP plot (right). For our future work, we can explore PP methods with other projection indices that are potentially more suitable for clustering. In particular, we will need to study how to execute the optimization of the projection indices and implement the methods, since many of them are not already implemented.

\begin{figure*}[htbp!]
    \centerline{\includegraphics[width=0.8\textwidth]{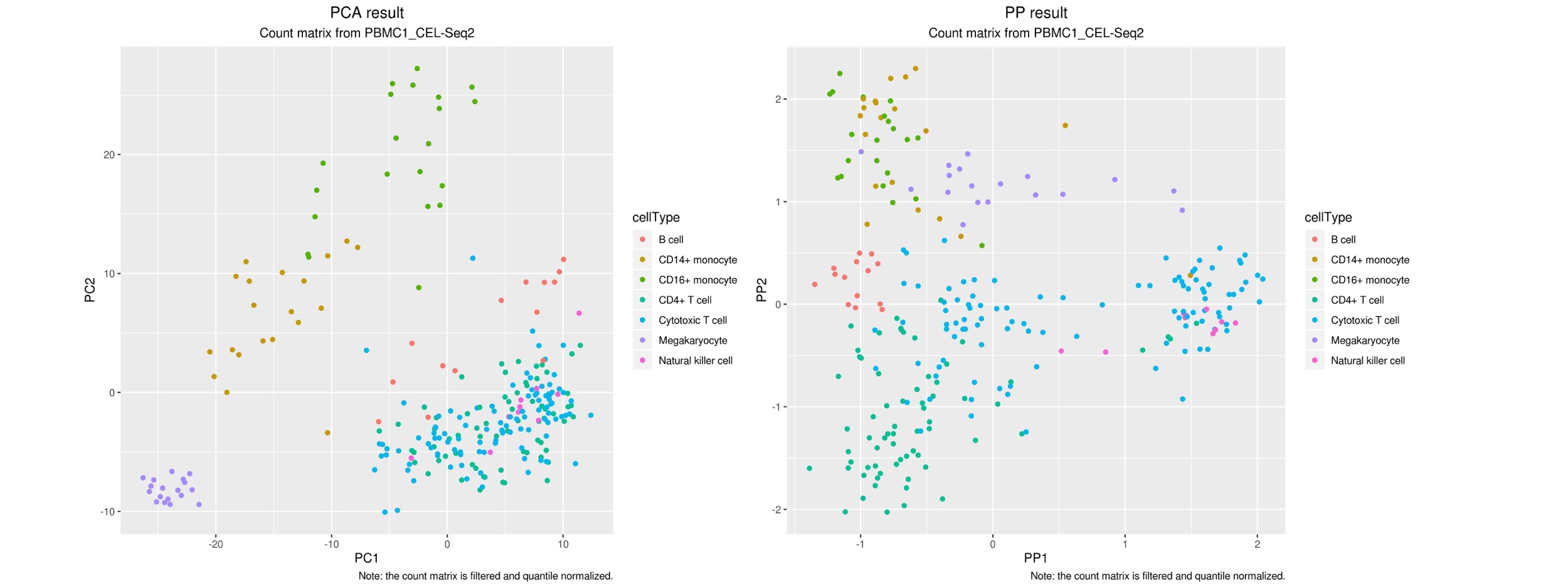}}
\end{figure*}

\begin{figure*}[htbp!]
    \vspace{-0.5in}
    \centerline{\includegraphics[width=0.8\textwidth]{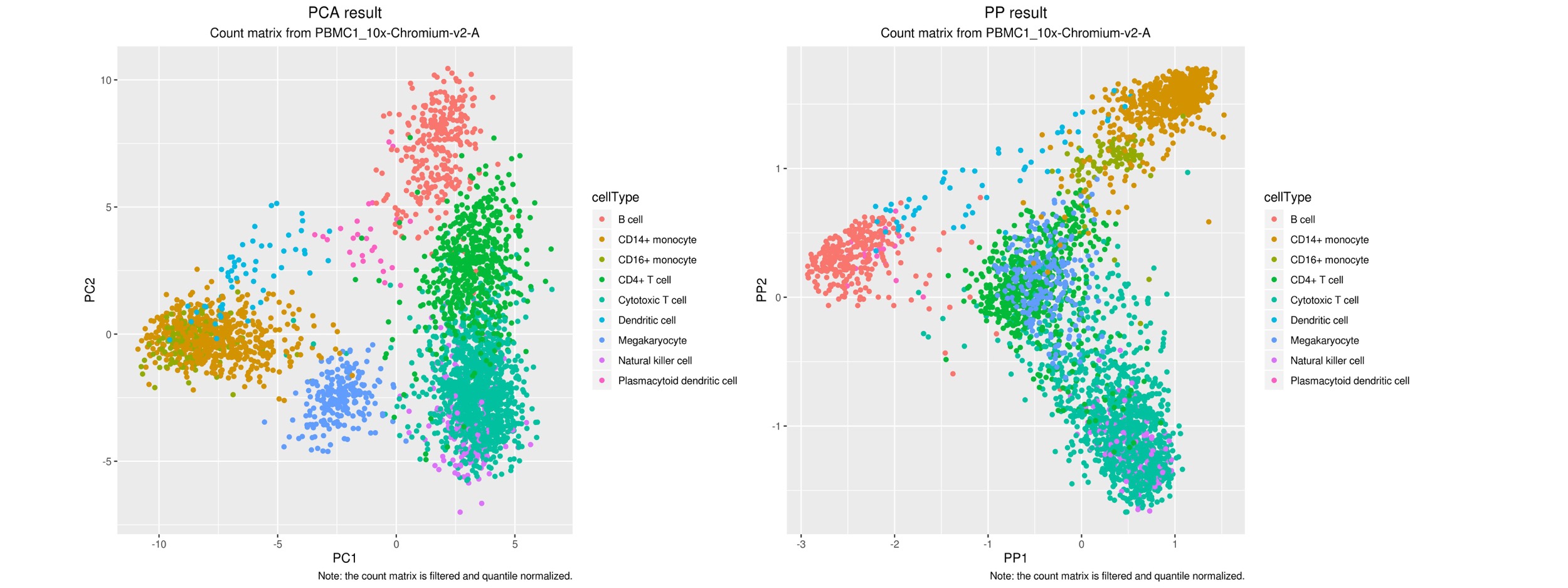}}
\end{figure*}

\begin{figure*}[htbp!]
    \vspace{-0.5in}
    \centerline{\includegraphics[width=0.8\textwidth]{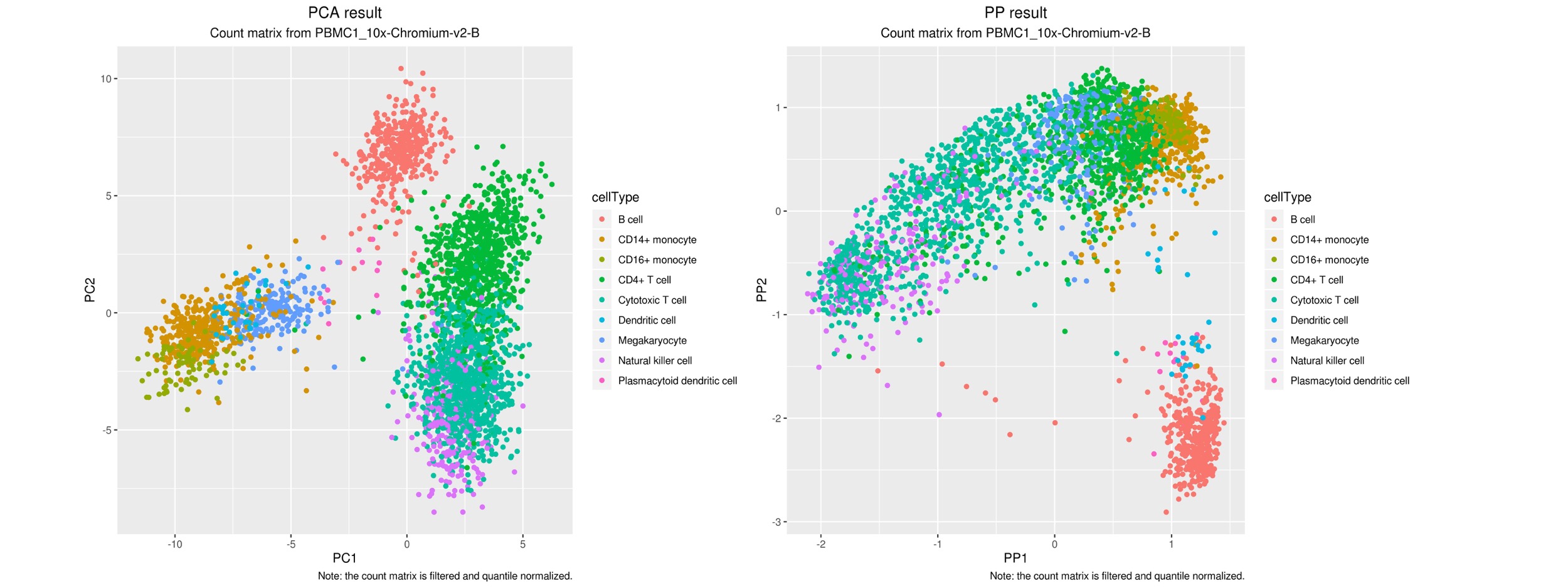}}
\end{figure*}

\begin{figure*}[htbp!]
    \vspace{-0.5in}
    \centerline{\includegraphics[width=0.8\textwidth]{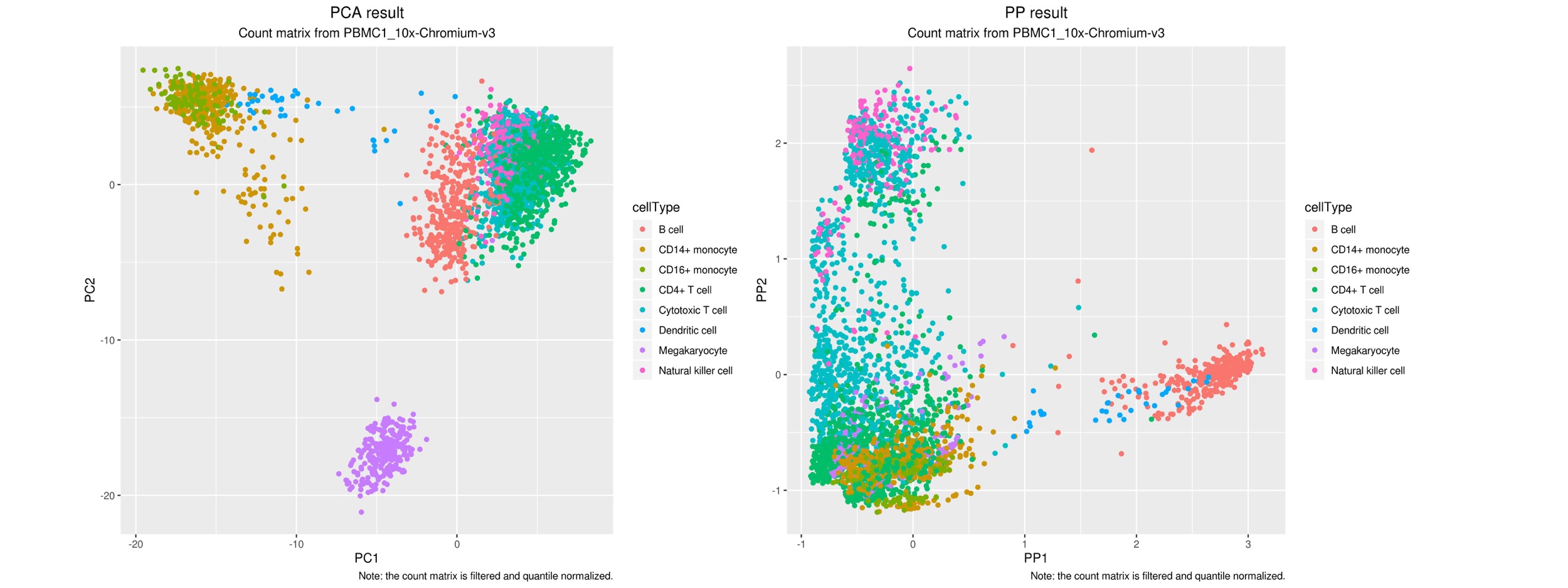}}
\end{figure*}

\begin{figure*}[htbp!]
    \vspace{-0.5in}
    \centerline{\includegraphics[width=0.8\textwidth]{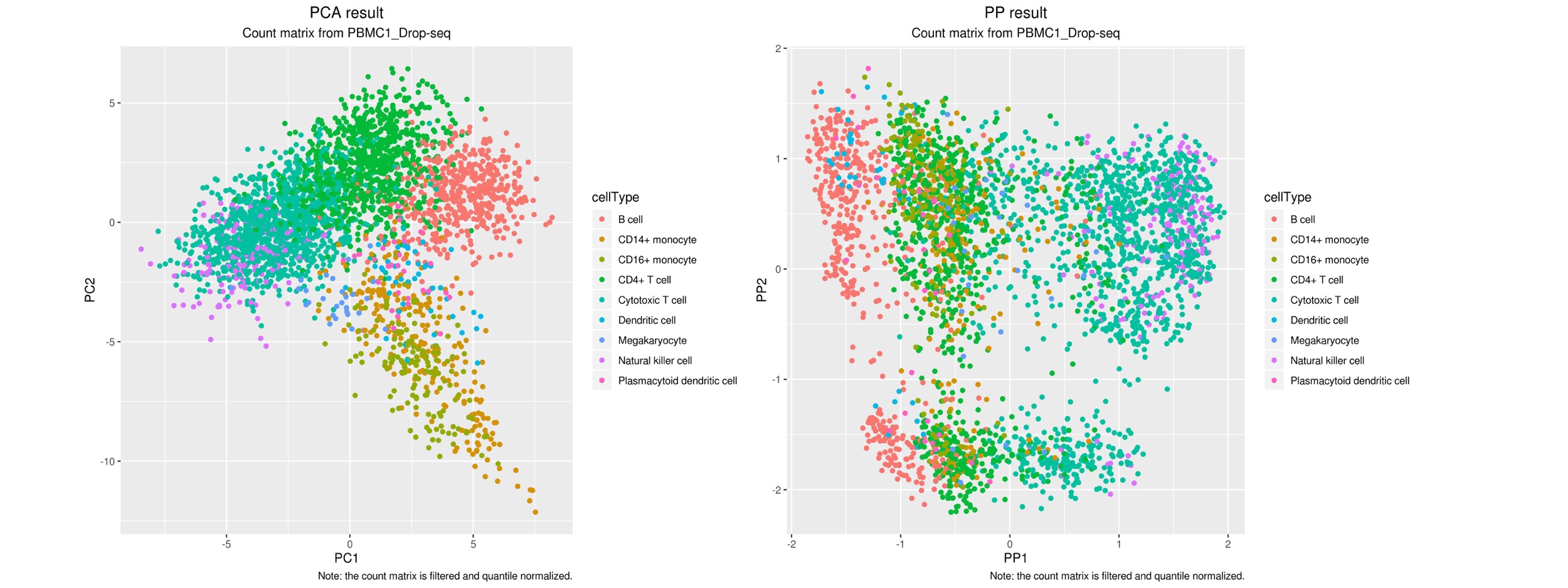}}
\end{figure*}

\begin{figure*}[htbp!]
    \vspace{-0.5in}
    \centerline{\includegraphics[width=0.8\textwidth]{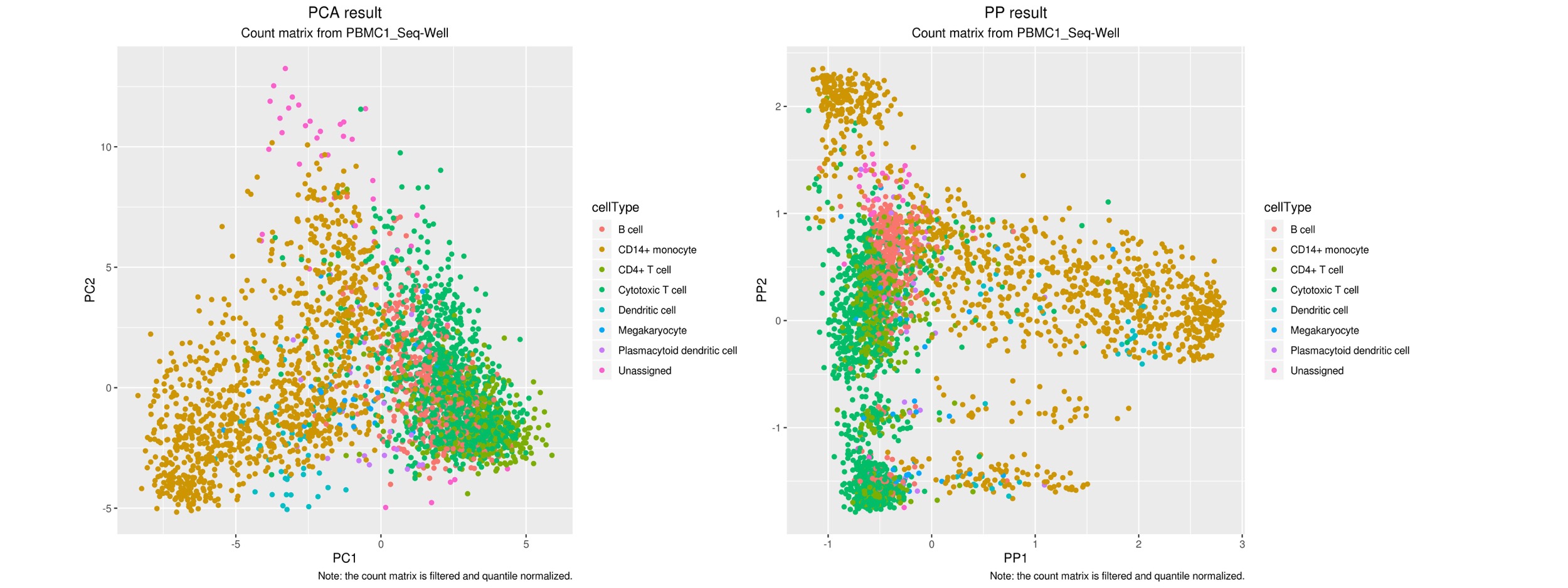}}
\end{figure*}

\bibliographystyle{unsrt}  
\bibliography{references}  

\newpage
\section{Appendix}
\begin{definition}[Entropy]
	The entropy of a probability measure $\mu$ dominated by a $\sigma$-finite counting measure on $\mathbb{R}^d$ is 
	$$H(\mu)=-\sum_{\omega\in\mathbb{R}^d}\mu(\{\omega\})\log\mu(\{\omega\})$$
\end{definition}
\begin{definition}[Differential entropy \cite{komogorov1956}]
	The differential entropy of a probability measure $\nu$ dominated by Lebesgue measure on $\mathbb{R}^d$ is defined as
	$$H(\nu)=-\int\frac{d\nu}{d\lambda_d}(\omega)\log\frac{d\nu}{d\lambda_d}(\omega)\lambda_k(d\omega)$$
	where $\lambda_d$ is the d-dim Lebesgue measure and $\frac{d\nu}{d\lambda_d}$ is the Radon-Nikodym derivative (i.e., see \cite{dabrowska2019}) of $\nu$ w.r.t. $\lambda_k$.
\end{definition}
\begin{definition}[Kullback–Leibler divergence]
	Let $\Prob$ and $\mathbb{Q}$ be equivalent probability measures dominated by a common measure $\mu$ (either counting or Lebesgue), then the Kullback–Leibler divergence (KL divergence) of $\Prob$ and $\mathbb{Q}$ is
	$$KL(\Prob||\mathbb{Q})=-\int \frac{d\Prob}{d\mu}\log(\frac{\frac{d\mathbb{Q}}{d\mu}}{\frac{d\Prob}{d\mu}})d\mu$$
\end{definition}

We have the following lemma of entropy.

\begin{lemma}[Entropy of affine transformation]\label{entropyaffine}
	Let $X$ be a random vector, then the entropy of $X$ is defined to be $H(X)=H(\Prob_X)$ where $\Prob_X$ is the induced probability measure. Let $Y=AX + b$, a non-singular affine transformation of $X$ ($A$ is invertible), then 
	$$H(Y)=\begin{cases}
	H(X) &\text{if $\Prob_X$ is discrete}\\
	H(X) + \log|\det A| &\text{if $\Prob_X\ll\lambda_d$}
	\end{cases}$$
\end{lemma}
The proof is based on change of variable formula in measure theory (i.e. theorem 6.3.1 in \cite{dabrowska2019}) and we have an important consequence:

\begin{lemma}[KL divergence is affine invariant]\label{KLdiv}
	Suppose $\Prob$ and $\mathbb{Q}$ are two probability measures dominated by counting or Lebesgue measure on $\mathbb{R}^d$ and $A\in\mathbb{R}^{d\times d}$ is invertible, then
	$$KL(\Prob||\mathbb{Q})=KL(\Prob_A||\mathbb{Q}_A)$$
	where $\Prob_A$ and $\mathbb{Q}_A$ are probability measures induced by the linear transformation $A$.
\end{lemma}

\end{document}